\documentclass[11pt]{article}
\usepackage[utf8]{inputenc}
\usepackage{newpxtext}
\usepackage[letterpaper, margin=1.3in]{geometry}
\usepackage[edtable]{lineno}
\usepackage{bm}
\usepackage{natbib}
\usepackage{amsmath}
\usepackage{graphicx}
\usepackage{booktabs}
\usepackage[noblocks]{authblk}
\usepackage[final]{pdfpages}
\usepackage{caption}
\usepackage{setspace}%for double spacing
\usepackage{standalone}%for including vignettes
\usepackage[hide]{todo}%for revisions
\usepackage[nolists,tablesfirst]{endfloat}%for revised submission
\usepackage{tcolorbox}%for reviewer replies
\usepackage[colorlinks = true,
            linkcolor = black,
            urlcolor  = blue,
            citecolor = black]{hyperref}%for links to commits in reviewer response
\newcounter{comments}[section]

\newcommand{\R}[1]{}
\newcommand{\lr}[1]{}
\usepackage{appendix}

\newcommand{\Rname}{\texttt{deBInfer}}

\usepackage{color}
\usepackage{fancyvrb}

\DefineVerbatimEnvironment{Highlighting}{Verbatim}{commandchars=\\\{\}}
\usepackage{framed}
\definecolor{shadecolor}{RGB}{248,248,248}

\newcommand{\KeywordTok}[1]{\textcolor[rgb]{0.13,0.29,0.53}{\textbf{{#1}}}}
\newcommand{\DataTypeTok}[1]{\textcolor[rgb]{0.13,0.29,0.53}{{#1}}}
\newcommand{\DecValTok}[1]{\textcolor[rgb]{0.00,0.00,0.81}{{#1}}}

\newcommand{\FloatTok}[1]{\textcolor[rgb]{0.00,0.00,0.81}{{#1}}}

\newcommand{\StringTok}[1]{\textcolor[rgb]{0.31,0.60,0.02}{{#1}}}

\newcommand{\CommentTok}[1]{\textcolor[rgb]{0.56,0.35,0.01}{\textit{{#1}}}}

\newcommand{\OtherTok}[1]{\textcolor[rgb]{0.56,0.35,0.01}{{#1}}}

\newcommand{\NormalTok}[1]{{#1}}
\usepackage{graphicx,grffile}
\makeatletter
\def\maxwidth{\ifdim\Gin@nat@width>\linewidth\linewidth\else\Gin@nat@width\fi}
\def\maxheight{\ifdim\Gin@nat@height>\textheight\textheight\else\Gin@nat@height\fi}
\makeatother
\title{\vspace{-3cm}\Rname: Bayesian inference for dynamical models of biological systems in R}
\author[1,2,3,*]{Philipp H. Boersch-Supan}
\affil[1]{Department of Integrative Biology, University of South Florida, Tampa, FL 33610}
\affil[2]{Emerging Pathogens Institute, University of Florida, Gainesville, FL 32610}
\affil[3]{Department of Geography, University of Florida, Gainesville, FL 32611}
\affil[4]{Department of Statistics, Virginia Polytechnic Institute and State University, Blacksburg, VA  24061}
\affil[*]{pboesu@gmail.com}
\author[2,3]{Sadie J. Ryan}
\author[1,4]{Leah R. Johnson}
\date{\today}
\begin{document}

\maketitle
\doublespacing

\begin{abstract}
\begin{enumerate}
 \item Understanding the mechanisms underlying biological systems, and ultimately, predicting their behaviours in a changing environment requires overcoming the gap between mathematical models and experimental or observational data. Differential equations (DEs) are commonly used to model the temporal evolution of biological systems, but statistical methods for comparing DE models to data and for parameter inference are relatively poorly developed.
       This is especially problematic in the context of biological systems where observations are often noisy and only a small number of time points may be available.
 \item The Bayesian approach offers a coherent framework for parameter inference that can account for multiple sources of uncertainty, while making use of prior information.
       It offers a rigorous methodology for parameter inference, as well as modelling the link between unobservable model states and parameters, and observable quantities. 
 \item We present \Rname, a package for the statistical computing environment R, implementing a Bayesian framework for parameter inference in DEs. \Rname\  provides templates for the DE model, the observation model and data likelihood, and the model parameters and their prior distributions. A Markov chain Monte Carlo (MCMC) procedure processes these inputs to estimate the posterior distributions of the parameters and any derived quantities, including the model trajectories. 
 Further functionality is provided to facilitate MCMC diagnostics, the visualisation of the posterior distributions of model parameters and trajectories, and the use of compiled DE models for improved computational performance\R{corr:ae4a}. 
 \item The templating approach makes \Rname\ applicable to a wide range of DE models. We demonstrate its application to ordinary and delay DE models for population ecology. 
\end{enumerate}

\end{abstract}


\section{Introduction}

The use of differential equations (DEs) to model dynamical systems has a long and fruitful tradition in biological disciplines such as epidemiology, population ecology, and physiology \citep{volterra1926fluctuations,kermack1927contribution}. 
As DE models are used in an attempt to understand biological systems, it is becoming clear that the simplest models cannot capture the rich variety of dynamics observed in them \citep{evans2013simple}.
However, more complex models come at the expense of additional states and/or parameters, and require more information for parameterization.
Further, as most observational datasets contain uncertainty, model identification and fitting become increasingly difficult \citep{lonergan2014data}.
Keeping complex models tractable and testable, and linking modeled quantities to data thus requires statistical methods of similar sophistication. 
This is particularly relevant in biology, where data series are often short or noisy, and where the scope for observational or experimental replication may be limited.

A vast array of analytical and numerical methods exists for solving DE models as well as exploring their properties and the effect of parameter values on their dynamics \citep{jones2003differential,smith2011an}. 
In some cases, parameters may be derived from first principles or measured directly, but often some or all parameters cannot be determined by either approach, and it is necessary to estimate them from an observational dataset.

Parameter estimation methods for DE models, and their implementation as computational tools, are still less well developed than the aforementioned system dynamics tools, and are a topic of active research. \R{corr:ae1}

Traditional parameter inference, also known as ``model calibration'' or ``solving inverse problems'', has, generally, been based on the maximum likelihood principle \citep{brewer2008fitting,aster2011parameter}, which assumes the existence of a true model $\mathcal{M}_{true}$ giving rise to a true dataset $\mathcal{Y}_{true}$ such that 

\begin{equation}\mathcal{M}_{true}(\boldsymbol\theta) = \mathcal{Y}_{true},\end{equation}
where $\boldsymbol\theta$ is the parameter set for the model\R{corr:re1-2}. The additional assumption that the observations $\mathcal{Y}$ arise from a sum of $\mathcal{Y}_{true}$ and measurement noise that is independently and normally distributed then leads to the least squares solution that is found by minimizing the Euclidian norm of the residual, 

\begin{equation}||\mathcal{M}(\theta)-\mathcal{Y}||_2.\end{equation} 
This approach has been applied to both ordinary differential equations (ODEs) \citep[e.g.][]{baker2005computational}, and simple delay-differential equations (DDEs) \citep[e.g.][]{horbelt2002parameter}. It allows for point estimates of the parameters, as well as the estimation of normal confidence intervals for the parameters and the correlations between them. 
However, these error bounds are local in nature and thus offer limited insight into the variability that is to be expected in the model outputs.

Bayesian approaches for parameter estimation in complex, nonlinear models were established early on \citep[e.g.][]{tarantola1982inverse,poole2000inference} and they are being applied with increasing frequency to a broad range of biological models \citep[e.g.][]{coelho2011bayesian,voyles2012temperature,johnson2013bayesian,smith2015inferred}. 
Recent methodological advances have included the application of Hamiltonian Monte Carlo to ODE models, realised in the software package Stan \citep{stan}, particle MCMC methods \citep{andrieu2010particle}, approximate Bayesian computation \citep[ABC; e.g.][]{liu2001combined,toni2009approximate}, and so called "plug-and-play" approaches \citep[e.g.][]{he2009plug}. 
A suite of these methods are implemented in the R package \texttt{pomp} \citep{king2016pomp}\R{corr:re2-2}.
While many statistical approaches, including the one presented here, treat the numerical solution of the DE model as exact, there has also been work towards quantifying the uncertainty contained in the numerical DE solutions themselves \citep{chkrebtii2013bayesian}.\R{corr:ae1b}

In the Bayesian approach the model, its parameters, and the data are viewed as random variables. 
This approach to parameter inference is attractive, as it provides a coherent framework that allows the incorporation of uncertainty in the observations and the process, and it relaxes the assumption of normal errors. 
It provides us not only with full probability distributions describing the parameters, but also with probability distributions for any quantity derived from them, including the model trajectories.
Further, the Bayesian framework naturally lets us incorporate prior information about the parameter values. This is particularly useful when there are known biological or theoretical constraints on parameters. For example, many biological parameters, such as body size cannot take on negative values. Using informative priors can help constrain the parameter space of the estimation procedure, aiding with parameter identifiability.

We explain the rationale behind the Bayesian approach below and describe our implementation of a fitting routine based on a Markov chain Monte Carlo (MCMC) sampler coupled to a numerical DE solver.
We illustrate the application of \verb+deBInfer+ to a simple example, the logistic differential equation, as well as a more complex model of the reproductive life history of the fungal pathogen \emph{Batrachochytrium dendrobatidis}.

\section{Methods}
The purpose of \Rname\ is to estimate the probability distribution of the parameters of a user specified DE model $\mathcal{M}$, given an empirical dataset $\mathcal{Y}$, and accounting for the uncertainty in the data.
The model takes the general form

\begin{equation}
 \mathcal{M} \equiv \frac{d\boldsymbol x}{dt}=\boldsymbol f(\boldsymbol x_t, t, \boldsymbol\theta)
\end{equation}
where $\boldsymbol x$ is a vector of variables evolving with time; $\boldsymbol f$ is a functional operator that takes a time input and a vector of continuous functions $\boldsymbol x_t(\theta)$ and generates the vector $\frac{d\boldsymbol x}{dt}$ as output; and $\boldsymbol \theta$ denotes a set of parameters. 
Further, we define $x_t(\boldsymbol\tau) = \boldsymbol x(t + \boldsymbol\tau)$. When all $\tau \in \boldsymbol\tau = 0$  the model is represented by a system of ODEs, when any $\tau < 0$ the model is represented by a system of delay-differential equations (DDEs). 
For the purposes of inference  $\boldsymbol \tau$ is simply a subset of the parameters $\boldsymbol \theta$ that are to be estimated. \Rname\ implements inference for ODEs as well as DDEs with constant delays.

Using Bayes's Theorem \citep{clark2007models} we can calculate the posterior distribution of the model parameters, given the data and the prior information as

\begin{equation}
\Pr(\boldsymbol\theta | \mathcal{Y}) = \frac{\Pr(\mathcal{Y}|\boldsymbol\theta) \Pr(\boldsymbol\theta)}{\int \Pr(\mathcal{Y}|\boldsymbol\theta)\Pr(\boldsymbol\theta) d\boldsymbol\theta} 
\end{equation}
where $\Pr()$ denotes a probability, $\mathcal{Y}$ denotes the data, and $\boldsymbol\theta$ denotes the set of model parameters. 
The product in the numerator is the  \emph{joint distribution}, which is made up of the \emph{likelihood} $\Pr(\mathcal{Y}|\boldsymbol\theta)$ or $\mathcal{L}(\mathcal{Y}|\boldsymbol\theta)$, which gives the probability of observing $\mathcal{Y}$ given the deterministic model $\mathcal{M}(\boldsymbol\theta)$, and the \emph{prior distribution} $\Pr(\boldsymbol\theta)$, which represents the knowledge about $\boldsymbol\theta$ before the data were collected.
The denominator represents the marginal distribution of the data $\Pr(\mathcal{Y})=\int \Pr(\mathcal{Y}|\boldsymbol\theta)\Pr(\boldsymbol\theta) d\boldsymbol\theta$. 
Before the data are collected $\mathcal{Y}$ is a random variable, but after they are collected the marginal distribution becomes a fixed quantity. 
This means, the inferential problem reduces to

\begin{equation}
 \Pr( \boldsymbol\theta | \mathcal{Y}) \propto \Pr(\mathcal{Y}|\boldsymbol\theta)\Pr(\boldsymbol\theta).
\end{equation}
That is, finding a specific proportionality that allows the \emph{posterior} $\Pr( \boldsymbol\theta | \mathcal{Y})$ to be a proper probability density (or mass) function that integrates to 1.

Closed form solutions for the posterior are practically impossible to obtain for complex non-linear models with more than a few parameters, but they can be approximated, e.g. by combining the MCMC algorithm with a Metropolis-Hastings sampler \citep{clark2007models}. This yields a sequence of likelihoods that follow a frequency distribution which approximates the posterior distribution.

The likelihood $\mathcal{L}(\mathcal{Y}|\boldsymbol\theta)$ describes the probability of the data for a given realization of the model $\mathcal{M}(\boldsymbol\theta)$, and we can use the fact that the data are uncertain to derive an expression like

\begin{equation}
 \mathcal{L}(\mathcal{Y}|\boldsymbol\theta) = \prod_t \mathcal{P}(\mathcal{Y}_{t},\mu=\mathcal{Y}_{t}(\theta),\sigma^2=\mathcal{V}_{t})
\end{equation}

where $\mathcal{P}$ is a parametric probability distribution, typically with first and second moments $\mu$ and $\sigma^2$,  $\mathcal{Y}_t$ is data item $t$, and $\mathcal{V}_t$ is the variance associated with $\mathcal{Y}_t$.

Often the data $\mathcal{Y}$ contain multiple data series, e.g. time-course observations of different state variables, following different probability distributions. 
In this case the likelihood becomes the product over all series and each data item in each series $s$\R{corr:re1-3} 

\begin{equation}
\label{eq:lik-multi}
 \mathcal{L}(\mathcal{Y}|\boldsymbol\theta) = \prod_s \prod_t \mathcal{P}_s(\mathcal{Y}_{s,t},\mu_s=\mathcal{Y}_{s,t}(\theta),\sigma^2_s=\mathcal{V}_{s,t}).
\end{equation}

\section{Implementation}
\Rname\ provides a framework for dynamical models consisting of a deterministic DE model and a stochastic observation model. 
In order to perform inference using \Rname, the user must specify R functions or data structures representing\R{corr:ae4b}: the DE model; an observation model, and thus the data likelihood; and declare all model and observation parameters, including prior distributions for those parameters that are to be estimated. The DE model itself can also be provided as a shared object, e.g. a compiled C function.
\Rname\ takes these inputs and performs MCMC to sample from the posterior distributions of parameters, solving the DE model numerically within the MCMC procedure. The MCMC procedure for \Rname\ offers independent, as well as random-walk Metropolis-Hastings updates and is implemented fully in R \citep{baseR}. Background on Metropolis-Hastings MCMC are widely available in the literature \citep[e.g.][]{clark2007models,brooks2011handbook}. 

\begin{table}[ht]
\small
\caption{ Implementation of the random-walk Metropolis-Hastings algorithm. The transition from a parameter value $\theta^{(k)}$ in the Markov chain at step $k$ to its value at step $k + 1$ proceeds via the outlined steps. $q$ is a conditional density, the so called \emph{proposal distribution}.}
\label{tab:rw-mcmc}
\begin{tabular}{p\textwidth}
\hline
\begin{enumerate}
\item Generate a proposal $\theta^{(*)} \sim q(\theta^{(*)}|\theta^{(k)})$
\item Evaluate the prior probability $\Pr(\theta^{(*)})$
\item \textbf{if} $\Pr(\theta^{(*)})=0$
\begin{description}
\item[] Let $\theta^{(k+1)}\leftarrow\theta^{(k)}$
\end{description}
\item  \textbf{if} $\Pr(\theta^{(*)})\neq 0$
\begin{description}
\item[]\textbf{if} $\theta \in \boldsymbol\theta_{DE}$: solve the DE model
\item[] Let
$
\theta^{(k+1)} \leftarrow \begin{cases}
\theta^{(*)} & \hbox{{ with probability} }\ \rho(\theta^{(k)},\theta^{(*)}), \cr
\theta^{(k)} & \hbox{{ with probability} }\ 1 - \rho(\theta^{(k)},\theta^{(*)}), \cr
\end{cases}
$\newline
{where}
$
\rho(\theta^{(k)},\theta^{(*)}) = \min \left\{ \dfrac{\Pr(\boldsymbol\theta^{(*)}|\mathcal{Y})}{\Pr(\boldsymbol\theta^{(k)}|\mathcal{Y})} \;
\dfrac{q(\theta^{(k)}|\theta^{(*)})}{q(\theta^{(*)}|\theta^{(k)})} \;, 1 \right\}
$
\end{description}
\end{enumerate}\\
\hline
\end{tabular}
\end{table}

As numerically solving the DE model is the most computationally costly step, we made two slight modifications to the basic Metropolis-Hastings algorithms. (i) \Rname\ makes a distinction between the parameters of the DE model $\boldsymbol\theta_{DE}$, and the observation parameters $\boldsymbol\theta_{obs}$, invoking the solver only for updates of the former, and (ii) the prior probability of each parameter proposal from the random walk sampler is evaluated before the posterior density and the acceptance ratio are calculated. This allows the rejection of proposals outside the prior support without invoking the numerical solver. The algorithm is outlined in Table~\ref{tab:rw-mcmc}.

\Rname\ provides a choice of three proposal distributions $q$ for the first step in the algorithm, a normal $\mathcal{N}(\theta^{(k)}, \sigma^2_{prop})$, an asymmetric uniform $\mathcal{U}(\frac{a}{b}\theta^{(k)}, \frac{b}{a}\theta^{(k)})$ and a multivariate normal $\mathcal{N}(\boldsymbol\theta^{(k)}, \boldsymbol\Sigma)$. \Rname\ requires manual tuning, i.e. the variance components $\sigma^2_{prop}$, $a$ and $b$, and $\boldsymbol\Sigma$, respectively, are user specified inputs. The asymmetric uniform distribution is useful for proposals of parameters that are strictly positive, such as variances, and the multivariate normal is useful for efficiently sampling parameters that are strongly correlated, as is often the case for DE model parameters.

\section{A simple example - logistic population growth}

\begin{table}
\caption{An overview of the main functions available in \Rname .}
\label{tab:funcs}
\small
\begin{tabular}{lp{0.65\textwidth}}
\hline
Function & Description\\
\hline
{\tt debinfer\_par} & creates a data structure representing an individual parameter or initial value of the DE model, or an observation parameter, and the corresponding values, priors, etc.\\
{\tt setup\_debinfer} & combines multiple parameter declarations into an input object for inference\\
{\tt de\_mcmc} & conducts MCMC inference on a DE model and returns an object of the class {\tt debinfer\_result}.\\
{\tt plot.debinfer\_result} & Plots traces and posterior densities (wrapper for {\tt coda::plot.mcmc}).\\
{\tt summary.debinfer\_result} & Summary statistics for MCMC samples (wrapper for {\tt coda::summary.mcmc}).\\
{\tt pairs.debinfer\_result} & Pairwise plots and correlations of marginal posterior distributions.\\
{\tt post\_prior\_densplot} & Overlay of posterior and prior densities for free parameters.\\
{\tt post\_sim}& Simulate posterior trajectories of the DE model and summary statistics thereof.\\
{\tt plot.post\_sim\_list}& Plot posterior DE model trajectories.\\

\hline
\end{tabular}
\normalsize
\end{table}

We illustrate the steps needed to perform inference for a DE model, by conducting inference on the logistic model (acknowledging that the existence of a closed form solution to this DE makes this an artificial example): 

\begin{equation}\frac{dN}{dt}=rN\left(1-\frac{N}{K}\right).\end{equation}
Annotated code to implement this model, simulate observations from it, and conduct the inference is provided as a package vignette (Appendix~\ref{app:logistic}).  An overview of the core functions available in \Rname\ is provided in Table~\ref{tab:funcs}.

\subsection{Installation}\label{preliminaries}

The deBInfer package is available on CRAN. The development version can be installed from github using \verb+devtools+ \citep{devtools}\R{corr:re1-6}, which can be installed\R{corr:ae2} from CRAN.

\begin{Highlighting}[]
\CommentTok{#Install the CRAN release.}
\KeywordTok{install.packages}\NormalTok{(}\StringTok{"deBInfer"}\NormalTok{)}
\CommentTok{#Alternatively install devtools and the development version of deBInfer.}
\KeywordTok{install.packages}\NormalTok{(}\StringTok{"devtools"}\NormalTok{)}
\NormalTok{devtools::}\KeywordTok{install_github}\NormalTok{(}\StringTok{"pboesu/debinfer"}\NormalTok{) }
\CommentTok{#Load deBInfer. }
\KeywordTok{library}\NormalTok{(deBInfer)}
\end{Highlighting}

\subsection{Specification of the differential equation model}
\Rname\ makes use of the \verb+deSolve+ and \verb+PBSddesolve+ packages \citep{soetaert2010solving,PBSddesolve} to numerically solve ODE and DDE models. 
The DE model has to be specified as a function containing the model equations, following the guidelines given in the respective package documentations. For our simple example the function takes three inputs: \verb+time+, a vector of time points at which to evaluate the DE, \verb+y+, a vector containing the initial value for the state variable $N$, and \verb+parms+, a vector containing the parameters $r$ and $K$.
\begin{Highlighting}[]
\NormalTok{logistic_model <-}\StringTok{ }\NormalTok{function(time, y, parms) \{ }
  \KeywordTok{with}\NormalTok{(}\KeywordTok{as.list}\NormalTok{(}\KeywordTok{c}\NormalTok{(y, parms)), \{}
    \NormalTok{dN <-}\StringTok{ }\NormalTok{r *}\StringTok{ }\NormalTok{N *}\StringTok{ }\NormalTok{(}\DecValTok{1} \NormalTok{-}\StringTok{ }\NormalTok{N /}\StringTok{ }\NormalTok{K) }
    \KeywordTok{list}\NormalTok{(dN)}
  \NormalTok{\})}
\NormalTok{\}}
\end{Highlighting}

\subsection{Observation model and likelihood specification}
For the purpose of demonstration we will conduct inference on simulated observations from this model %with the underlying parameters $r=0.1$, $K = 10$ and
assuming log-normal noise with a standard deviation $\sigma^2_{obs }$. A set of simulated observations is provided with the package and can be loaded  with the command {\tt data(logistic)}.
The appropriate log-likelihood takes the form 

\begin{equation}
\ell(\mathcal{Y}|\boldsymbol\theta) = \sum_t \ln\left(\frac{1}{\tilde{N_t}\sigma_{obs}\sqrt{2\pi}}\exp\left(-\frac{(\ln \tilde{N_t}-\ln (N_t+\varepsilon))^2}{2\sigma^2_{obs}}\right)\right)
\end{equation}
where $\tilde{N}_t$ are the observations, and $N_t$ are the predictions of the DE model given the current MCMC sample of the parameters $\boldsymbol{\theta}$. Further, $\varepsilon \ll 1$ \R{corr:re1-5} is a small correction needed, because the exact DE solution can equal zero (or less, depending on numerical precision of the solver). $\varepsilon$ should therefore be at least as large as the expected numerical precision of the solver.  We chose $\varepsilon= 10^{-6}$, which is on the same order as the default numerical precision of the default solver (\texttt{deSolve::ode} with \texttt{method = "lsoda"}), but we found that the inference results were insensitive to this choice as long as $\varepsilon \leq 0.01$ (Appendix~\ref{app:logistic}, Section 7).

The \Rname\ observation model template requires three inputs: a {\tt data.frame} of observations, \verb+data+; the simulated trajectory returned by the numerical solver in MCMC procedure, \verb+sim.data+; and the current sample of the parameters, \verb+samp+. The user specifies the observation model such that it returns the summed log-likelihoods of the data. In this example the observations are in the data.frame column \verb+N_noisy+, and the corresponding predicted states are in the column \verb+N+ of the matrix-like object \verb+sim.data+ (see Appendix~\ref{app:logistic}).
\begin{Highlighting}[]
\CommentTok{#load example data}
\KeywordTok{data}\NormalTok{(logistic)}
\CommentTok{# user defined data likelihood}
\NormalTok{logistic_obs_model <-}\StringTok{ }\NormalTok{function(data, sim.data, samp)\{}
  \NormalTok{epsilon <-}\StringTok{ }\FloatTok{1e-6}
  \NormalTok{llik <- }\KeywordTok{sum}\NormalTok{(}\KeywordTok{dlnorm}\NormalTok{(data$N_noisy, }\DataTypeTok{meanlog = }\KeywordTok{log}\NormalTok{(sim.data[, }\StringTok{"N"}\NormalTok{]+ epsilon),}
                     \DataTypeTok{sdlog = }\NormalTok{samp[[}\StringTok{"sdlog.N"}\NormalTok{]],}\DataTypeTok{log = }\OtherTok{TRUE}\NormalTok{))}
  \KeywordTok{return}\NormalTok{(llik) }
  \NormalTok{\}}
\end{Highlighting}

\subsection{Parameter, prior, and sampler specification}
All parameters that are used in the DE model and the observation model need to be declared for the inference procedure using the \verb+debinfer_par()+ function. The declaration describes the variable name, whether it is a DE or observation parameter and whether or not it is to be estimated. If the parameter is to be estimated, the user also needs to specify a prior distribution and a number of additional parameters for the MCMC procedure. \verb+debinfer+ currently supports priors from all probability distributions implemented in base R, as well as their truncated variants, as implemented in the \verb+truncdist+ package \citep{truncdist}.

We declare the DE model parameter $r$, assign a prior $r \sim \mathcal{N}(0,1)$ and a random walk sampler with a Normal kernel (\verb+samp.type="rw"+) and  proposal variance of 0.005 with the command
\begin{Highlighting}[]
\NormalTok{r <-}\StringTok{ }\KeywordTok{debinfer_par}\NormalTok{(}\DataTypeTok{name =} \StringTok{"r"}\NormalTok{, }\DataTypeTok{var.type =} \StringTok{"de"}\NormalTok{, }\DataTypeTok{fixed =} \OtherTok{FALSE}\NormalTok{,}
                \DataTypeTok{value =} \FloatTok{0.5}\NormalTok{, }\DataTypeTok{prior = }\StringTok{"norm"}\NormalTok{, }\DataTypeTok{hypers = }\KeywordTok{list}\NormalTok{(}\DataTypeTok{mean =} \DecValTok{0}\NormalTok{, }\DataTypeTok{sd =} \DecValTok{1}\NormalTok{),}
                \DataTypeTok{prop.var = }\FloatTok{0.005}\NormalTok{, }\DataTypeTok{samp.type = }\StringTok{"rw"}\NormalTok{)}
\end{Highlighting}
Similarly, we declare $K\sim \ln\mathcal{N}(1,1)$ and $\sigma^2_{obs}\sim\ln\mathcal{N}(0,1)$.
\begin{Highlighting}[]
\NormalTok{K <-}\StringTok{ }\KeywordTok{debinfer_par}\NormalTok{(}\DataTypeTok{name =} \StringTok{"K"}\NormalTok{, }\DataTypeTok{var.type =} \StringTok{"de"}\NormalTok{, }\DataTypeTok{fixed =} \OtherTok{FALSE}\NormalTok{,}
                \DataTypeTok{value =} \DecValTok{5}\NormalTok{, }\DataTypeTok{prior = }\StringTok{"lnorm"}\NormalTok{, }\DataTypeTok{hypers = }\KeywordTok{list}\NormalTok{(}\DataTypeTok{meanlog =} \DecValTok{1}\NormalTok{, }\DataTypeTok{sdlog =} \DecValTok{1}\NormalTok{),}
                \DataTypeTok{prop.var = }\FloatTok{0.1}\NormalTok{, }\DataTypeTok{samp.type = }\StringTok{"rw"}\NormalTok{)}

\NormalTok{sdlog.N <-}\StringTok{ }\KeywordTok{debinfer_par}\NormalTok{(}\DataTypeTok{name =} \StringTok{"sdlog.N"}\NormalTok{, }\DataTypeTok{var.type =} \StringTok{"obs"}\NormalTok{, }\DataTypeTok{fixed =} \OtherTok{FALSE}\NormalTok{,}
                \DataTypeTok{value =} \FloatTok{0.1}\NormalTok{, }\DataTypeTok{prior = }\StringTok{"lnorm"}\NormalTok{, }\DataTypeTok{hypers = }\KeywordTok{list}\NormalTok{(}\DataTypeTok{meanlog =} \DecValTok{0}\NormalTok{, }\DataTypeTok{sdlog =} \DecValTok{1}\NormalTok{),}
                \DataTypeTok{prop.var = }\KeywordTok{c}\NormalTok{(}\DecValTok{3}\NormalTok{,}\DecValTok{4}\NormalTok{), }\DataTypeTok{samp.type = }\StringTok{"rw-unif"}\NormalTok{)}
\end{Highlighting}
Note that we are using the asymmetric uniform proposal distribution for the variance parameter (\verb+samp.type="rw-unif"+), as this ensures strictly positive proposals.
Lastly, we provide an initial value $N_0=0.1$ for the DE:
\begin{Highlighting}[]
\NormalTok{N <-}\StringTok{ }\KeywordTok{debinfer_par}\NormalTok{(}\DataTypeTok{name =} \StringTok{"N"}\NormalTok{, }\DataTypeTok{var.type =} \StringTok{"init"}\NormalTok{, }\DataTypeTok{fixed =} \OtherTok{TRUE}\NormalTok{, }\DataTypeTok{value =} \FloatTok{0.1}\NormalTok{)}
\end{Highlighting}

\subsection{MCMC inference}
The MCMC procedure is called using the function \verb+de_mcmc()+ which takes the declared parameters, the DE and observational models, the data, and further optional arguments to the MCMC procedure and/or the solver as inputs and returns an array containing the resulting MCMC samples.

All declared parameters are collated using \verb+setup_debinfer()+

\begin{Highlighting}[]
\NormalTok{mcmc.pars <-}\StringTok{ }\KeywordTok{setup_debinfer}\NormalTok{(r, K, sdlog.N, N)}
\end{Highlighting}
and passed to \verb+de_mcmc()+ which is set to use \verb+deSolve::ode()+ as a backend in this case, as specified by the argument \verb+solver = "ode"+
\begin{Highlighting}[]
\CommentTok{# do inference with deBInfer}
\CommentTok{# MCMC iterations}
\NormalTok{iter <-}\StringTok{ }\DecValTok{5000}
\CommentTok{# inference call}
\NormalTok{mcmc_samples <-}\StringTok{ }\KeywordTok{de_mcmc}\NormalTok{(}\DataTypeTok{N =} \NormalTok{iter, }\DataTypeTok{data = }\NormalTok{logistic, }\DataTypeTok{de.model = }\NormalTok{logistic_model, }
                          \DataTypeTok{obs.model = }\NormalTok{logistic_obs_model, }\DataTypeTok{all.params = }\NormalTok{mcmc.pars,}
                          \DataTypeTok{Tmax =} \KeywordTok{max}\NormalTok{(logistic$time), }\DataTypeTok{data.times = }\NormalTok{logistic$time, }
                          \DataTypeTok{cnt = }\DecValTok{500}\NormalTok{, }\DataTypeTok{plot = }\OtherTok{FALSE}\NormalTok{, }\DataTypeTok{solver = }\StringTok{"ode"}\NormalTok{)}
\end{Highlighting}

\subsection{Inference Outputs}
\begin{figure}[tbp]
\centering
 \includegraphics[width=0.5\textwidth]{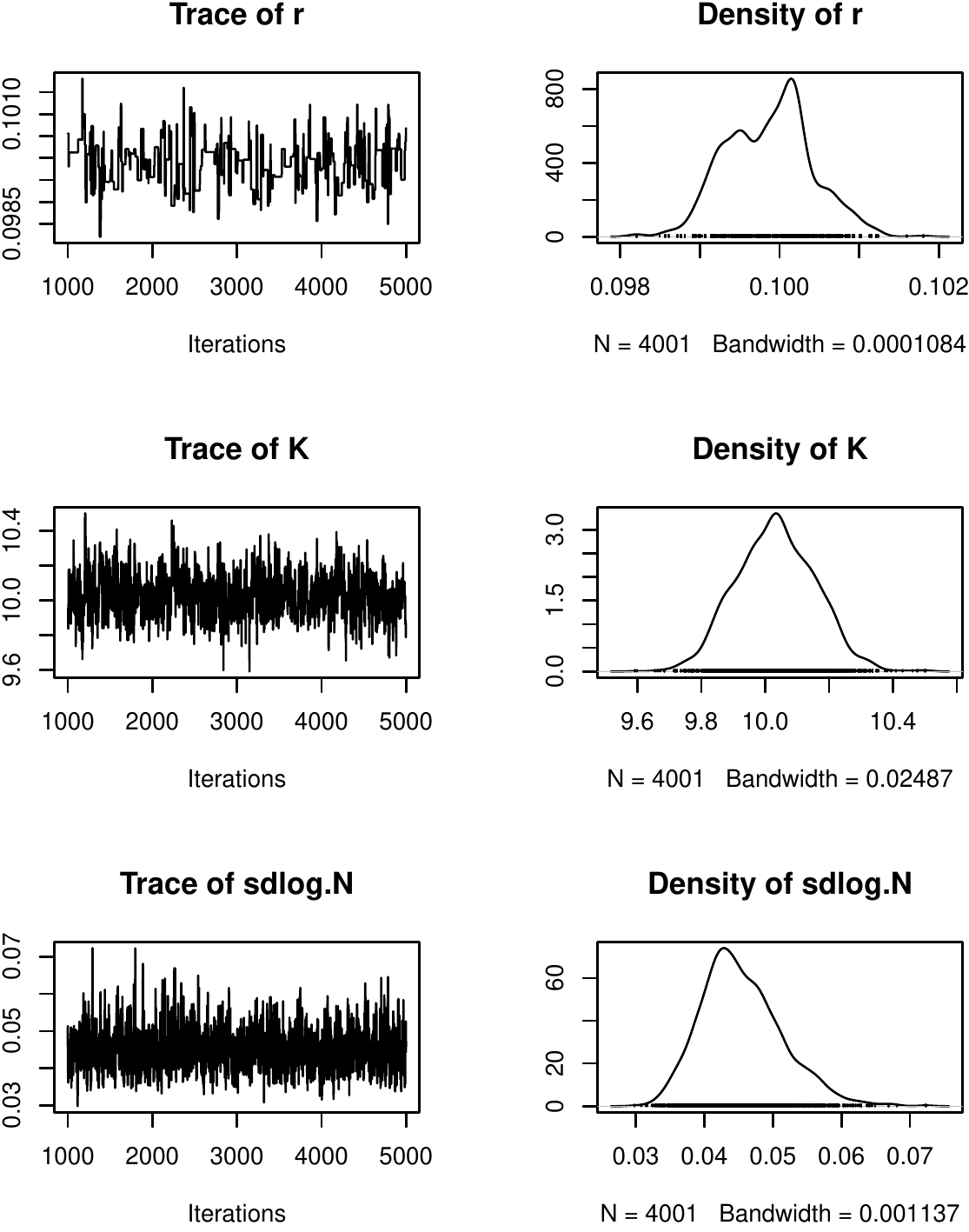}
 \caption{MCMC traces and posterior density plots for the logistic model. Figures like this one can be created using \texttt{plot.debinfer\_result}.}
 \label{fig:logistic-outputs-1}
\end{figure}

\begin{figure}[tbp]
\centering
 \includegraphics[width=0.5\textwidth]{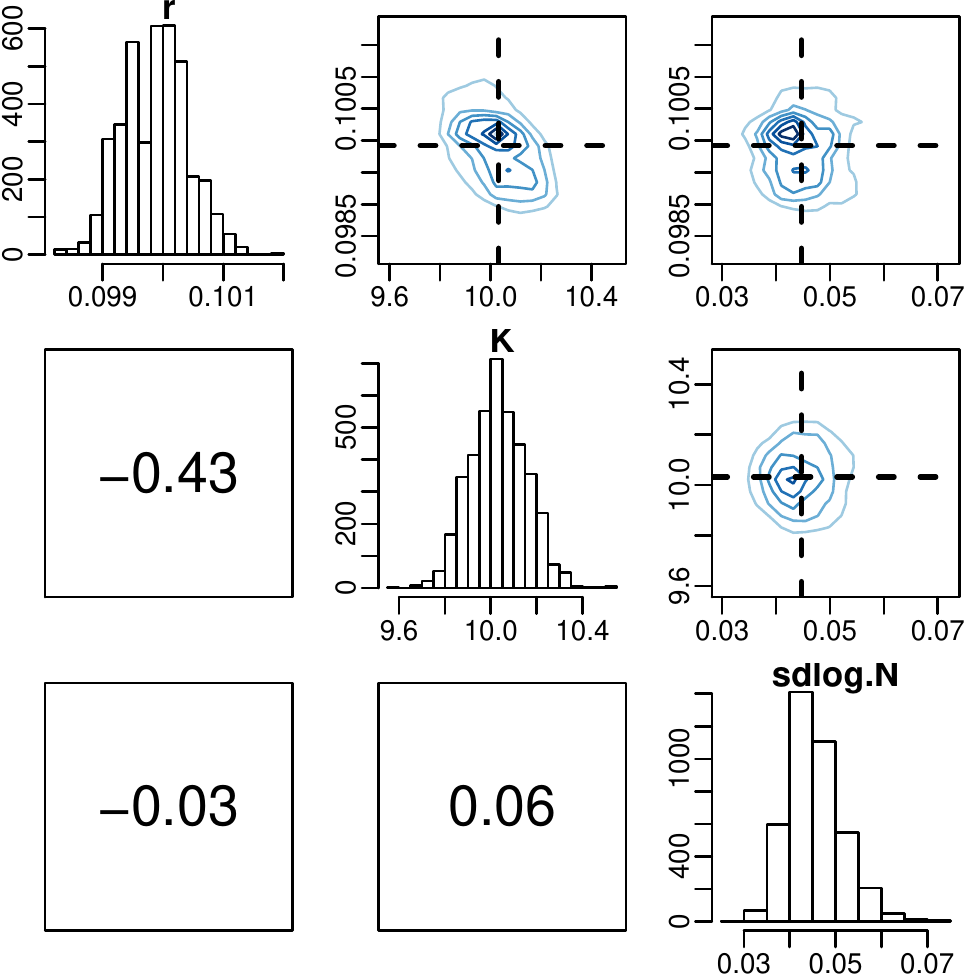}
 \caption{Pairwise plot of the marginal posterior distributions. This figure was created using \texttt{pairs.debinfer\_result}.}
 \label{fig:logistic-outputs-2}
\end{figure}

\begin{figure}[tbp]
\centering
 \includegraphics[width=0.8\textwidth]{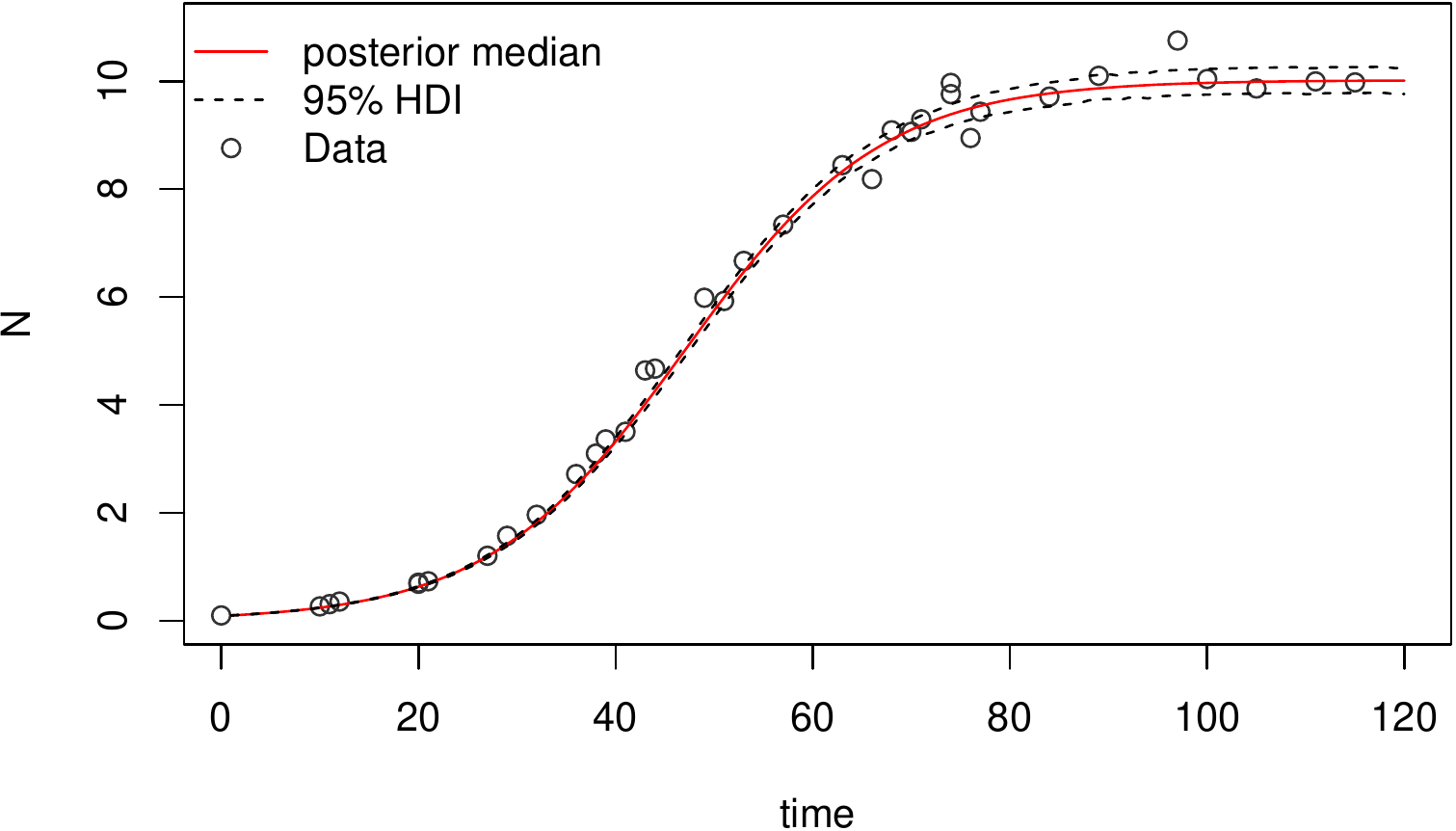}
 \caption{Posterior model trajectory (median with 95\% highest posterior density interval), created with \texttt{plot.post\_sim\_list}, and the data points used for fitting.}
 \label{fig:logistic-outputs-3}
\end{figure}

The inference function returns an object of class \verb+debinfer_result+, which contains the posterior samples in a format compatible with the \verb+coda+ package \citep{coda}, as well as the DE and observation models and all parameters used for inference. This allows the use of the diagnostic functions and plotting routines provided in \verb+coda+ (see Fig.~\ref{fig:logistic-outputs-1}). We also provide additional functions and methods such as \verb+pairs.debinfer_result()+ to create pairwise plots of the marginal posterior distributions, which show correlations between individual parameters (see Fig.~\ref{fig:logistic-outputs-2}), \verb+post_prior_densplot()+, which allows a visual comparison between prior and marginal posterior densities for each parameter, \verb+post_sim()+ which simulates posterior model trajectories and associated credible intervals, as well as plotting methods for the latter (see Fig.~\ref{fig:logistic-outputs-3}).

\section{Example application - DDE model of fungal population growth}
To illustrate applications of \Rname\ beyond the simplistic example above, we outline inference procedures for a more complex model and corresponding observational data. Full model details and annotated code can be found in Appendix~\ref{app:chytrid}.

Our example demonstrates parameter inference for a DDE model of population growth in the environmentally sensitive fungal pathogen \emph{Batrachochytrium dendrobatidis} (Bd), which causes the amphibian disease chytridiomycosis \citep{rosenblum2010deadly,voyles2012temperature}. This model has been used to further our understanding of pathogen responses to changing environmental conditions. Further details about the model development, and the experimental procedures yielding the data used for parameter inference can be found in \citet{voyles2012temperature}.

The model follows the dynamics of the concentration of an initial cohort of zoospores, $C$, the concentration of zoospore-producing sporangia, $S$, and the concentration of zoospores in the next generation $Z$. The initial cohort of zoospores, $C$, starts at a known concentration, and zoospores in this initial cohort settle and become sporangia at rate $s_r$, or die at rate $\mu_Z$. $f_s$ is the fraction of sporangia that survive to the zoospore-producing stage. We assume that it takes a minimum of $T_{min}$ days before the sporangia produce zoospores, after which they produce zoospores at rate $\eta$. Zoospore-producing sporangia die at rate $d_s$. The concentration of zoospores, $Z$, is the only state variable  measured in the experiments, and it is assumed that these zoospores settle ($s_r$) or die ($\mu_Z$) at the same rates as the initial cohort of zoospores.
%\subsubsection{DDE model}
The equations that describe the population dynamics are as follows:

\begin{align}
\frac{dC}{dt} &= -(s_r +\mu_Z) C(t) \\
\frac{dS}{dt} &= s_r f_s C(t - T_{min})-d_s S(t)\\
\frac{dZ}{dt} &= \eta S(t) - (s_r+\mu_Z) Z(t)
\end{align}

Because the observations are counts of zoospores (i.e. discrete numbers), we assume that observations of the system at a set of discrete times $t'$ are independent Poisson random variables with a mean given by the solution of the DDE, at times $t'$.

\begin{figure}[tbp]
 \includegraphics[width=0.9\textwidth]{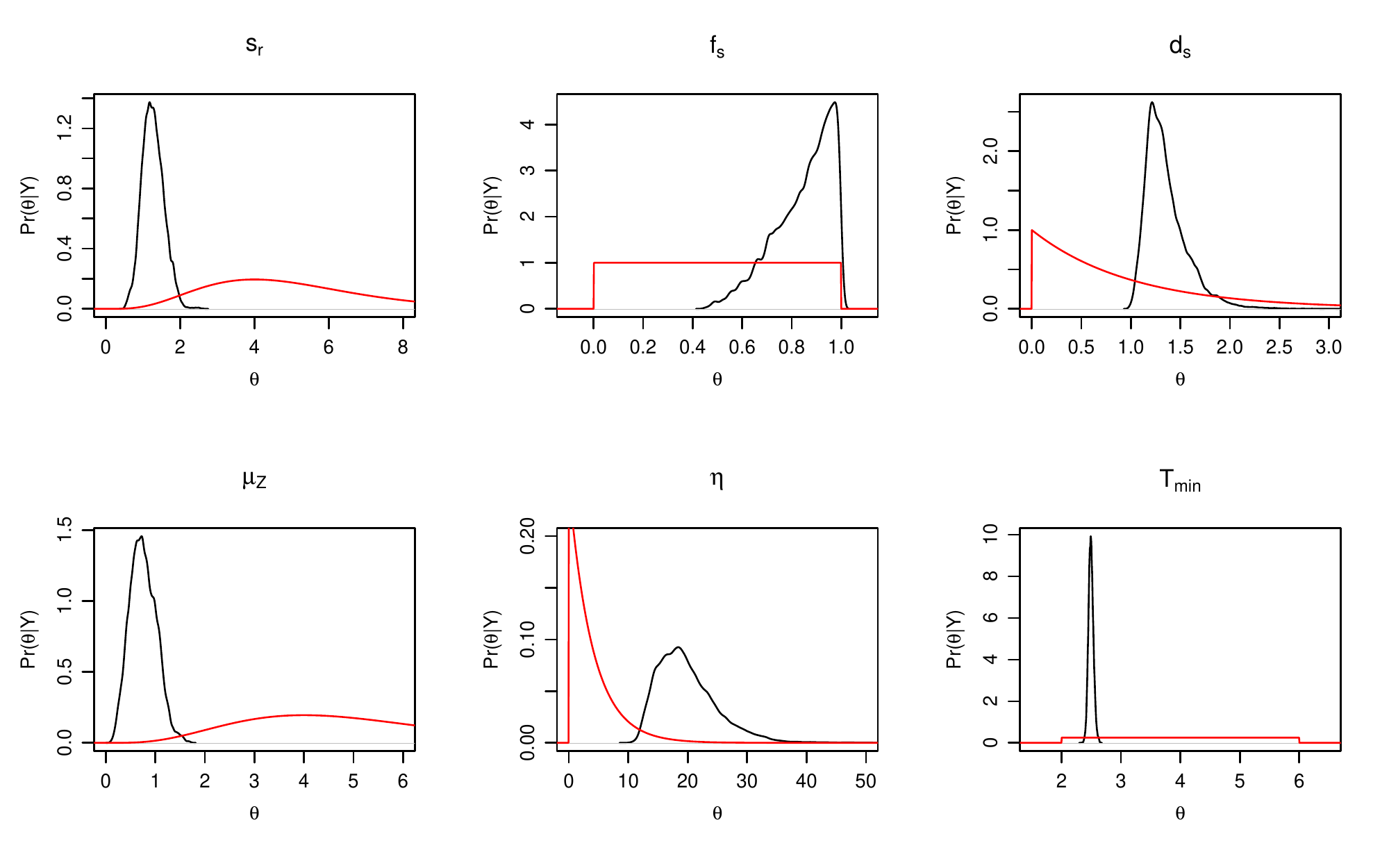}
 \caption{Comparison of marginal posterior densities (black) and the corresponding priors (red) of the estimated parameters of the chytrid model. This plot was created using {\tt post\_prior\_densplot}.}
 \label{fig:chytrid-outputs-2}
\end{figure}

\begin{figure}[tbp]
 \includegraphics[width=1.0\textwidth]{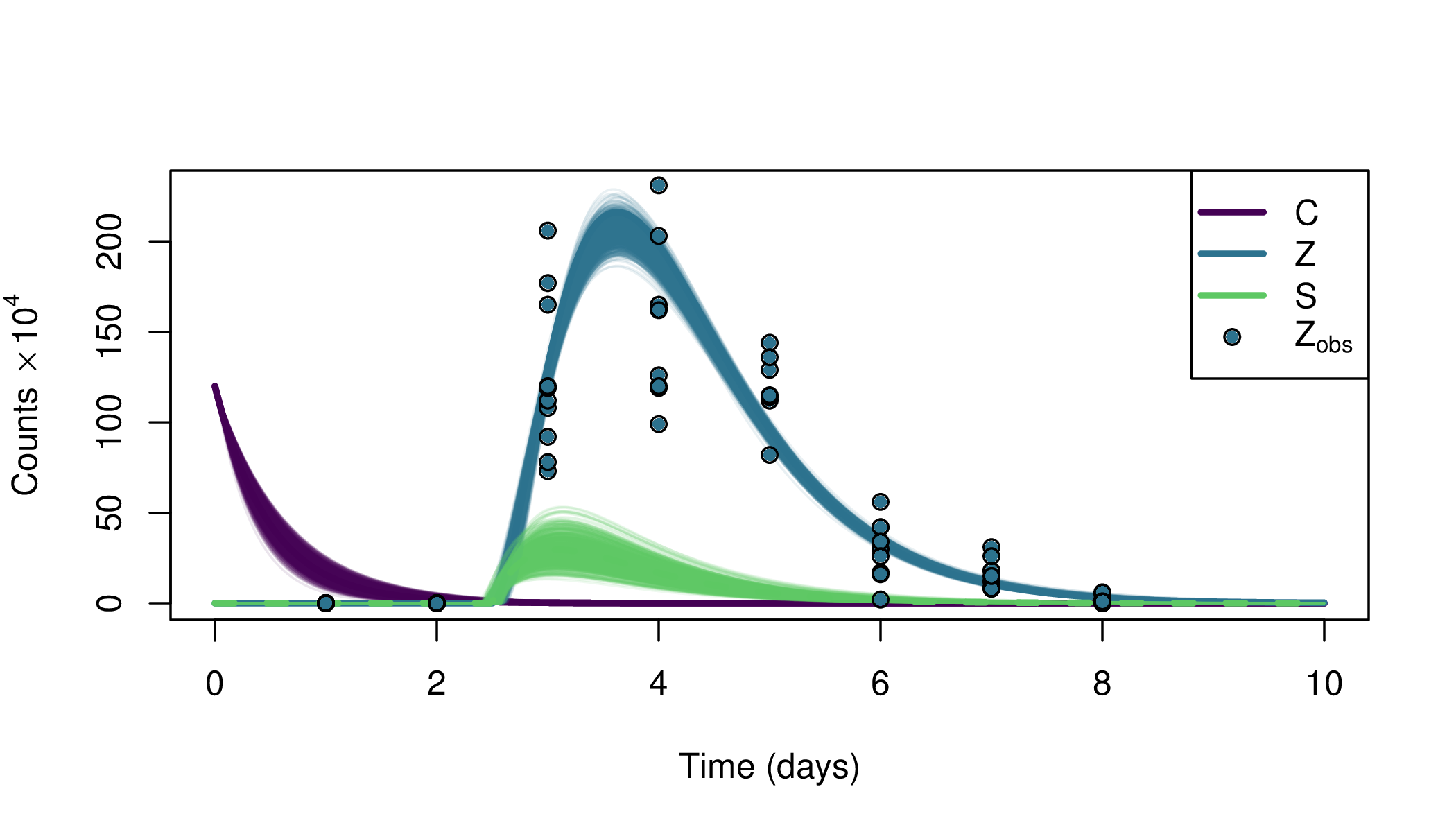}
 \caption{Posterior trajectories for each state variable of the chytrid model based on 1000 model simulations from the posterior of the parameters and the data points $Z_{obs}$ used for fitting.}
 \label{fig:chytrid-outputs}
\end{figure}

The log-likelihood of the data given the parameters, underlying model, and initial conditions is then a sum over the $n$ observations at each time point in $t'$

\begin{equation}
\ell(\mathbf{Z}|\mathbf{\theta})=\sum\limits^n_{t}Z_t \text{ log }\lambda-n\lambda
\end{equation}
%\subsubsection{Results}
In this case we conduct inference using \verb+deSolve::dede()+ as the backend to \verb+de_mcmc+. The marginal posteriors of the estimated parameters are presented in Fig.~\ref{fig:chytrid-outputs-2}, and posterior trajectories for the model are presented in Fig.~\ref{fig:chytrid-outputs}.

\section{Known limitations} 
The MCMC sampler is implemented in R, which makes it considerably slower than samplers written in compiled languages e.g., those underlying packages such as Stan \citep{stan} or Filzbach \citep{filzbach}\R{corr:re1-6b}. 
For inference conducted purely in R, the computational bottleneck is solving the DE model numerically. 
However, even for relatively simple models, a 5-10 fold speed up of the inference procedure can be achieved by using compiled DE models (see Appendix~\ref{app:compiled}) \R{corr:ae4c}.
Furthermore, the \verb+debinfer+ MCMC algorithm is not adaptive and requires manual tuning. 
Lastly, sampling using the Metropolis-Hastings MCMC algorithm itself can be inefficient in the presence of strong parameter correlations. Alternative approaches such as Hamiltonian MC \citep{stan} or particle-filtering methods \citep[e.g.][]{king2016pomp} may offer more efficient means for parameter estimation in ODEs in these cases.\R{corr:ae1c}
Nonetheless, the package is able to fit real world problems in a matter of minutes to hours on current desktop hardware, which is acceptable for many applications, while providing flexible inference for both ODE and DDE models\R{corr:re1-7}. 

\section{Conclusion}
Understanding the mechanisms underlying biological systems, and ultimately, predicting their behaviours in a changing environment requires overcoming the gap between mathematical models and experimental or observational data. We believe that Bayesian inference provides a powerful tool for fitting dynamical models and selecting between competing models. The \Rname\ R package provides a suite of tools to this end in a programming language that is widespread in many biological disciplines. We hope that our package, will lower the hurdle to the uptake of this inference approach for empirical biologists. We encourage users to report bugs and provide other feedback on the project issue page: https://github.com/\-pboesu/debinfer/issues

\section{Acknowledgements}
We thank Richard FitzJohn and two anonymous reviewers for their constructive comments on earlier versions of the code and manuscript.
All authors were supported by the U.S. National Science Foundation (Grant PLR-1341649). 
We thank Jamie Voyles for sharing the chytrid growth data.
The authors have no conflict of interest to declare.

\section{Data Accessibility}
All code and data used in this article are included in the \Rname\ package and its vignettes, which are freely available from CRAN: https://cran.r-project.org/package=deBInfer. The development version of the package is available at https://github.com/pboesu/debinfer.

\section{Author contributions}
LRJ conceived the methodology and wrote the initial R implementation; PHBS reimplemented the methodology as an R package; PHBS and SJR wrote the package documentation; PHBS led the writing of the manuscript. All authors tested the software, contributed critically to the drafts, and gave final approval for publication.

\bibliographystyle{mee}
\bibliography{debinfer}

%TC:ignore
\begin{appendices}
\section{Annotated code for the logistic DE example} \label{app:logistic}
This appendix can be found in the supplementary materials. It can also be displayed after installing \Rname\ with the R command:\\
{\tt vignette("logistic\_ode\_example", package="deBInfer")}

\section{Annotated code for the DDE example} \label{app:chytrid}
This appendix can be found in the supplementary materials. It can also be displayed after installing \Rname\ with the R command:\\
{\tt vignette("chytrid\_dede\_example", package="deBInfer")}

\section{Inference for a compiled DE model} \label{app:compiled}\R{corr:ae4d}
This appendix can be found in the supplementary materials. It can also be displayed after installing \Rname\ with the R command:\\
{\tt vignette("deBInfer\_compiled\_code", package="deBInfer")}

\end{appendices}

\processdelayedfloats

%TC:endignore
\end{document}